\documentclass[seceq]{iis}
\usepackage{graphics}
\usepackage{color}
\usepackage{amsmath}
\usepackage{caption}


\title{How to estimate the number of self-avoiding walks over $10^{100}$
?\\Use random walks.}
\runtitle{How to estimate the number of SAW over $10^{100}$? Use RW.}

\author{%
\name{Nobu C. \surname{Shirai}}$^{1,2}$
\CAE{shirai@cp.cmc.osaka-u.ac.jp} and
\name{Macoto  \surname{Kikuchi}}$^{2,1,3}$ 
}

\inst{$^{1}$ Graduate School of Science, Osaka University, \address{Toyonaka, Osaka 560-0043, Japan}\\
$^{2}$ Cybermedia Center, Osaka University, \address{Toyonaka, Osaka 560-0043, Japan}\\
$^{3}$ Graduate School of Frontier Biosciences, Osaka University, \address{Suita, Osaka 565-0871, Japan}}

\runauthor{Nobu C. Shirai}

\support{This work is supported by the Global COE Program Core Research and Engineering of Advanced Materials-Interdisciplinary Education Center for Materials Science, MEXT, Japan.}

\recdate{20YY}{0}{DD}
\accdate{20YY}{0}{DD}

\abst{
Counting the number of $N$-step self-avoiding walks (SAWs) on a lattice is one of the most difficult problems of enumerative combinatorics. Once we give up calculating the exact number of them, however, we have a chance to apply powerful computational methods of statistical mechanics to this problem. In this paper, we develop a statistical enumeration method for SAWs using the multicanonical Monte Carlo method.
A key part of this method is to expand the configuration space of SAWs to random walks, the exact number of which is known.
Using this method, we estimate a number of $N$-step SAWs on a square lattice, $c_N$, up to $N=256$. The value of $c_{256}$ is $5.6(1)\times 10^{108}$ (the number in the parentheses is the statistical error of the last digit) and this is larger than one googol ($10^{100}$).
}

\kword{\kw{Self-avoiding walks}, \kw{statistical enumeration}, \kw{multicanonical Monte Carlo method}, \kw{rare event sampling}}

\begin{document}
\maketitle
\section{Introduction}
\label{Introduction}
Starting from the origin of a lattice and connecting a line to one neighboring site after another, we get a path. If we choose a random direction for each step, the path becomes a random walk. The question we seek to answer is "How many $N$-step random walks are there on a square lattice?" Since there are four possible directions for each step, the answer is $4^N$. If we restrict the path by limiting visits to the same site, we get a self-avoiding walk (SAW) instead of a random walk.
We then ask the similar question. 
"How many $N$-step SAWs are there on a square lattice?" The answer, however, is not so simple as with random walks. The number of SAWs, $c_N$, is not known for large values of $N$.
There are some sophisticated enumeration methods for SAWs and exact values for $c_N$ on a square lattice are known only up to $N=71$~\cite{Jensen2004}.
The enumerated $c_N$ of $N=71$ is
\[
c_{71}=4\ 190\ 893\ 020\ 903\ 935\ 054\ 619\ 120\ 005\ 916
\sim 4.19\times 10^{30},
\]
which is much larger than Avogadro's constant.
In order to go further, we change the problem from calculating all digits of $c_N$ to estimating the first two or three digits of $c_N$. 
How large a $c_N$ can we estimate? 
Here, we set a tentative goal of a googol ($10^{100}$).

\begin{figure}[htb]
\centering
\begin{minipage}{72mm}
\includegraphics[width=55mm,keepaspectratio,clip]{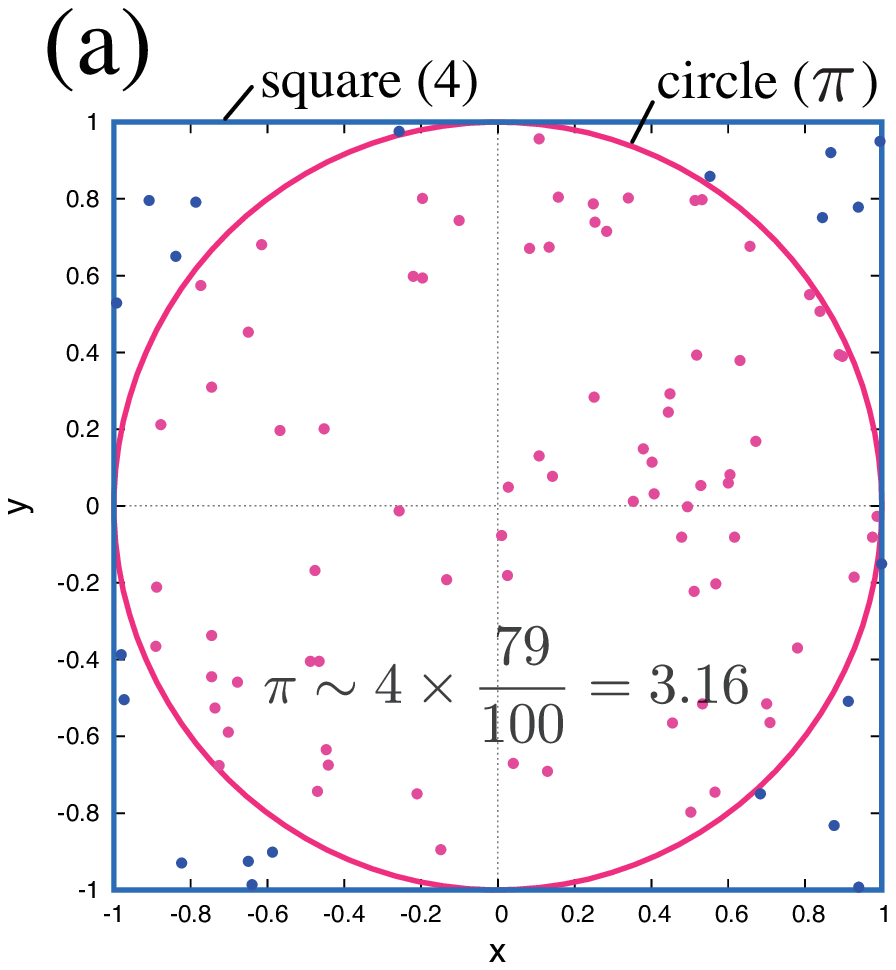}
\end{minipage}
\begin{minipage}{72mm}
\includegraphics[width=66mm,keepaspectratio,clip]{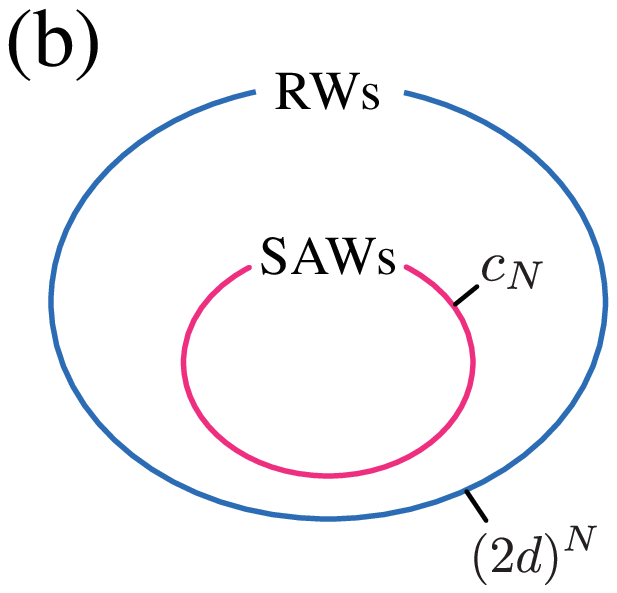}
\end{minipage} 
\caption{(a) Estimating the area of the unit circle using Monte Carlo integration. The values in the parentheses are the areas of the shapes. 79 of 100 points were plotted inside the circle (pink dots) and the area of the circle can be estimated as $4\times 79/100=3.16$. (b) A Venn diagram of SAWs and random walks. In order to calculate $c_N$, we expand the configuration space of SAWs to random walks, the number of which is exactly $(2d)^N$.}
\label{fig:MC_pi}
\end{figure}

One popular strategy in enumeration is {\it divide and conquer}.
We propose, however, a different strategy to estimate a large $c_N$, in which we expand the configuration space of SAWs to that of random walks and use knowledge of the number of random walks.
In order to explain how to use this exact number, we describe a statistical method of estimating the area of a unit circle using Monte Carlo integration as an example.
In Fig.~\ref{fig:MC_pi} (a), we show a unit circle surrounded by a square, the area of which is exactly 4.
To calculate the area of the circle, we randomly generate $n$ points on the square using pairs of real random numbers $(r_1,r_2)\ (-1\le r_1,r_2 \le 1)$ and count the number of pairs which satisfy the condition $x^2+y^2 \le 1$.
Let this number be $n_\mathrm{c}$.
Using $n_\mathrm{c}$ and $n$, we can estimate the area ratio of the circle to the square.
Since we know that the area of the square is 4, we can finally estimate the area of the unit circle as $4\times n_\mathrm{c}/n$. 
Let us apply this logic to the estimation of $c_N$.

In Fig.~\ref{fig:MC_pi} (b), we show a Venn diagram of SAWs and random walks.
The number of $N$-step random walks on a $d$-dimensional hypercubic lattice is exactly known as $(2d)^N$.
Instead of generating random points as in the above example, we use Markov chain Monte Carlo (MCMC) to calculate the ratio of the number of SAWs to that of random walks.
Once we obtain the ratio, we can estimate $c_N$ by multiplying the ratio by $(2d)^N$.

\section{The statistical enumeration method for self-avoiding walks}
\subsection{SAWs are rare random walks}
We define a self-avoiding walk (SAW) on a $d$-dimensional hypercubic lattice and its mathematical notation.
We denote an $N$-step path as $\omega=(\omega(0),\omega(1),\cdots,\omega(N))$, where each of $\omega(i)\ (i=0,1,\cdots,N)$ denotes a point which is randomly selected from the lattice.
When we impose two conditions, namely (i) $\omega(0)$ is at the origin of the lattice, and (ii) $|\omega(i+1)-\omega(i)|=1\ (i=0,1,\cdots,N-1)$ on $\omega$, $\omega$ becomes an $N$-step random walk.
If we add one more condition (iii) $\omega(i)\neq \omega(j)\ (i\neq j)$ on $\omega$, $\omega$ becomes an $N$-step SAW~\cite{MadrasSlade1993}.
It is obvious from the definition of a SAW that the set of $N$-step SAWs is a subset of that of $N$-step random walks.
SAWs are known as $n\to 0$ limit of $n$-vector model and have attracted a great interest of physicists in the context of critical phenomena.
As we mentioned above, we expand the configuration space of SAWs to that of random walks and try to sample SAWs using MCMC.

\begin{figure}[htb]
\centering
\includegraphics[width=50mm,keepaspectratio,clip]{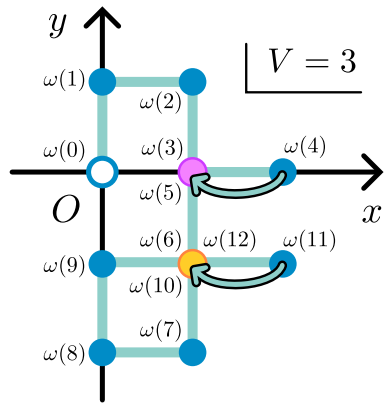}
\caption{A 12-step random walk with three intersections. 
There are one intersection on the purple circle and two intersections on the orange circle.
Total intersections $V$ of the path is 3.}
\label{fig:def_saw}
\end{figure}

Before we move on to the sampling method, we would like to discuss how rare SAWs are. The asymptotic behavior of $c_N$ is conjectured to be
\begin{equation}
c_N = A\mu^N N^{\gamma-1}[1+o(1)],
\end{equation}
where $\mu$ is called the connective constant, which is known to be $\sim 2.64$~\cite{Jensen2004}.
$\gamma$ is a critical exponent, which is thought to be universal. Namely, it depends only on the spatial dimensionality and it is independent of the lattice type.
In order to roughly estimate the ratio of the number of SAWs to that of random walks on a square lattice, we use the approximation
\begin{equation}
\frac{c_N}{(2d)^N} \sim \left( \frac{\mu}{4} \right)^N \sim \left(\frac{2.64}{4} \right)^N.
\end{equation}
In the case of $N=50$, $150$ and $250$, the ratios are
\[
\left( \frac{2.64}{4} \right)^{50} \sim 10^{-9}, 
\left( \frac{2.64}{4} \right)^{150} \sim 10^{-27}, 
\left( \frac{2.64}{4} \right)^{250} \sim 10^{-45}.
\]

Even with $N=50$, SAWs are very rare, and if we randomly generate random walks, we need about $10^9$ samples to get one SAW.
In the case of $N=250$, SAWs become extremely rare and we cannot get one SAW by random sampling even if we use the massive computer resources available today. 
Thus, we need a more efficient sampling method.

\subsection{Dividing the configuration space of random walks by the number of intersections}
\begin{figure}[tb]
\centering
\includegraphics[width=125mm,keepaspectratio,clip]{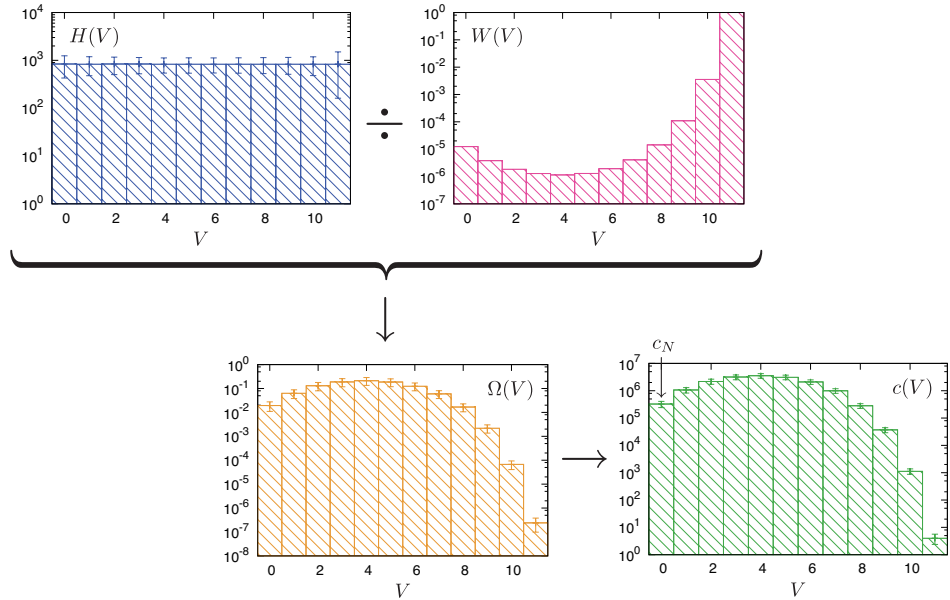}
\caption{A procedure of calculating $c(V)$. $c(V)$ is the number of random walks with $V$ intersections, and is calculated by $c(V)=(2d)^N \Omega(V)/\sum^{N-1}_{V=0} \Omega(V)$.}
\label{fig:cV_making}
\end{figure}
As a first step towards an efficient sampling method, we introduce the number of intersections $V$ as a parameter.
In Fig.~\ref{fig:def_saw}, we show a configuration of a random walk with three intersections, as an example.
If $k$ points of a path $(k\ge 3)$ are on the same site, we define that there are $k-1$ intersections on that site.
We divide the configuration space into $N$ parts ($V=0,1,2,\cdots,N-1$) according to the value of $V$, and configurations with $V=0$ are all SAWs.
If we introduce the energy of random walks by $V$, we get the modified Domb-Joyce model~\cite{DombJoyce1972,ShiraiKikuchi2012}.
In the high-and low-temperature limits, this model reduces to random walks and SAWs respectively.
From the discussion above, we need the ratio of the number of SAWs to that of random walks to estimate $c_N$. If we denote the density of states of the modified Domb-Joyce model as $\Omega(V)$, $\Omega(0)/\sum^{N-1}_{V=0} \Omega(V)$ gives the desired ratio, and we can calculate $c_N$ by
\begin{equation}
c_N = (2d)^N \frac{\Omega(0)}{\sum^{N-1}_{V=0}\Omega(V)}. \label{eq:c_N}
\end{equation}
In order to calculate $c_N$ accurately, we should ensure the accuracy of not only $\Omega(0)$ but $\Omega(V)$ for the entire range of $V$. 
It is difficult to do so, however, because the value of $\Omega(V)$ varies across many orders of magnitude.
To overcome these difficulty, we use a powerful computational method of statistical mechanics.

\subsection{Multicanonical simulation of the modified Domb-Joyce model}
In the next step, we explain how to sample SAWs. 
We use the multicanonical Monte Carlo method~\cite{BergNeuhaus1991,BergNeuhaus1992} to accurately estimate the density of states of the modified Domb-Joyce model, $\Omega(V)$, over a wide range of $V$.
This calculation is essentially the same as that of the multi-self-overlap ensemble~\cite{IbaKikuchi1998,ChikenjiIba1999} without Hamiltonian.
A key to this method is the weight function $W(V)$ that is built up to be proportional roughly to $1/\Omega(V)$ by iterative methods. 
Frequently-used iterative methods are as follows:
\begin{itemize}
\item the multicanonical method~\cite{BergNeuhaus1991,BergNeuhaus1992}
\item the entropic sampling method~\cite{Lee1993}
\item the Wang-Landau method~\cite{WangLandau2001,WangLandau2001PRE}.
\end{itemize}
We used the Wang-Landau method.
We omit explaining these iterative methods here, and explain the multicanonical method, assuming that $W(V)$ has already been obtained.
In the multicanonical method, the transition probability from one path, $\omega_a\ (V=V_a)$, to another path $\omega_b\ (V=V_b)$ is given by
\begin{equation}
p(\omega_a \to \omega_b) = 
\min\left[
\frac{W(V_b)}{W(V_a)},1
\right], \label{eq:transition_prob}
\end{equation}
where this form of the transition probability is the same as that of the Metropolis method with $W(V)$ instead of the Boltzmann factor.
An MCMC simulation using this transition probability generates an ensemble called a multicanonical ensemble. 
Sampling from this ensemble gives a flat histogram $H(V)$ (in the top left graph of Fig.~\ref{fig:cV_making}).
Dividing $H(V)$ by $W(V)$, we obtain $\Omega(V)$ (in the bottom left graph of Fig.~\ref{fig:cV_making}). 
From Eq.~\ref{eq:c_N} with $\Omega(V)$ obtained, we can finally calculate $c_N$ (in the bottom right graph of Fig.~\ref{fig:cV_making}).

\section{Results}
Using the proposed method above, we calculate $c_N$ with a statistical error up to $N=256$.
Some of the calculated $c_N$ values are shown in Table~\ref{table:c_N2D}, together with the available exact values $(N\le 71)$ and estimated values given by Prellberg and Krawczyk using flatPERM~\cite{PrellbergKrawczyk2004}.
The errors were estimated from the standard errors of histograms.
The estimated values agree well with exact values and estimated values of flatPERM.
We should note that flatPERM achieved longer walks up to $c_{1028}=1.74\times 10^{438}$, but due to the lack of error estimation, it is now known how many digits of the values are reliable.
The largest value of our estimation is
\[
c_{256}=5.6(1)\times 10^{108},
\]
and we achieved the goal of a $c_N$ value of over $10^{100}$.

\begin{table}[htb]
\caption{List of exact and estimated $c_N$ values on a square lattice \label{table:c_N2D}}
\begin{center}
\begin{tabular}{rrrr}
\hline\hline
$N$ & \multicolumn{1}{c}{$c_N$ (exact)} & \multicolumn{1}{c}{$c_N$ (flatPERM\cite{PrellbergKrawczyk2004})} & \multicolumn{1}{c}{$c_N$ (our method)} \\
\hline
4 & $100$ & & $1.0001(2)\times 10^{2}$\\
8 & $5916$ & & $5.915(2)\times 10^{3}$\\
16 & $1.7245\times 10^{7}$ & $1.7265\times 10^{7}$ & $1.724(1)\times 10^{7}$\\
32 & $1.1903\times 10^{14}$ & $1.1911\times 10^{14}$ & $1.186(2)\times 10^{14}$\\
64 & $4.5493\times 10^{27}$ & $4.5529\times 10^{27}$ & $4.54(2)\times 10^{27}$\\
128 & & $5.2970\times 10^{54}$ & $5.32(6)\times 10^{54}$\\
256 & & $5.6700\times 10^{108}$ & $5.6(1)\times 10^{108}$\\
\hline\hline
\end{tabular}
\end{center}
\end{table}

\section*{Acknowledgments}
We would like to thank Dr. T. Prellberg for provideing us detailed data of $c_N$ given by flatPERM.
This work was supported by the Global COE Program Core Research and Engineering of Advanced Materials-Interdisciplinary Education Center for Materials Science, MEXT, Japan

\end{document}